**The Western Africa Ebola virus disease epidemic exhibits both global exponential and local polynomial growth rates**

Gerardo Chowell[1,2,3], Cécile Viboud[2], James M. Hyman[4], Lone Simonsen[2,5]

[1] School of Public Health, Georgia State University, Atlanta, Georgia, USA

[2] Division of International Epidemiology and Population Studies, Fogarty International Center, National Institutes of Health, Bethesda, MD, USA

[3] School of Human Evolution and Social Change, Arizona State University, Tempe, AZ, USA

[4] Department of Mathematics, Tulane University, New Orleans, USA

[5] Department of Global Health, School of Public Health and Health Services, George Washington University, Washington DC, USA

## Abstract

**Background:**

While many infectious disease epidemics are initially characterized by an exponential growth in time, we show that district-level Ebola virus disease (EVD) outbreaks in West Africa follow slower polynomial-based growth kinetics over several generations of the disease.

**Methods:**

We analyzed epidemic growth patterns at three different spatial scales (regional, national, and subnational) of the Ebola virus disease epidemic in Guinea, Sierra Leone and Liberia by compiling publicly available weekly time series of reported EVD case numbers from the patient database available from the World Health Organization website for the period 05-Jan to 17-Dec 2014.

**Results:**

We found significant differences in the growth patterns of EVD cases at the scale of the country, district, and other subnational administrative divisions. The national cumulative curves of EVD cases in Guinea, Sierra Leone, and Liberia show periods of approximate exponential growth. In contrast, local epidemics are asynchronous and exhibit slow growth patterns during 3 or more EVD generations, which can be better approximated by a polynomial than an exponential.

**Conclusions:**

The slower than expected growth pattern of local EVD outbreaks could result from a variety of factors, including behavior changes, success of control interventions, or intrinsic features of the disease such as a high level of clustering. Quantifying the contribution of each of these factors could help refine estimates of final epidemic size and the relative impact of different mitigation efforts in current and future EVD outbreaks.

**Introduction**

A strain of Zaire Ebola virus emerged in West Africa in approximately December 2013 and, as of December 2014, has continued to spread extensively through Guinea, Sierra Leone, and Liberia, with limited transmission chains occurring in neighboring countries (Mali, Nigeria, and Senegal), and sporadic importations elsewhere. A total of 19497 Ebola virus disease (EVD) cases, with 7588 deaths, have been reported to the World Health Organization as of December 24, 2014 [1]. Halting the spread continues to be a challenging task as the virus is moving through high population density areas. Although the causative strain is closely related to a strain associated with past EVD outbreaks in Central and East Africa [2], the ongoing epidemic in West Africa was largely facilitated by delays in the implementation of control interventions and a lack of public health infrastructure.

Although past EVD outbreaks have been limited in geographic scope to rural areas and have been contained within just a few disease generations [3], the current epidemic has greater geographic scope, duration, morbidity, and mortality impact than ever before. Although unfortunate, the prolonged epidemic in West Africa offers a unique window to study the modes of spatial diffusion of EVD through a wide geographic region. A better understanding of the transmission dynamics of the ongoing EVD epidemic could improve our ability to forecast the final size and geographic scope of the epidemic and guide future mitigation efforts.

The cumulative number of EVD cases in West Africa grew exponentially during the early stages of the epidemic when the whole region or individual countries are considered (Guinea, Sierra Leone, and Liberia) [4-13]. Here we study spatially disaggregated data and show that local outbreaks are asynchronous and that they follow a slower growth pattern that can be best approximated by polynomial rather than exponential functions. These results contrast with the early growth of many infectious diseases, such as influenza, which is well approximated by exponential growth [14,15]. We then discuss potential mechanisms that could explain the slower than expected growth dynamics of EVD spread, including the natural history of the disease, spatial heterogeneity of the infected population, behavior changes, and control interventions.

**Data and Methods**

To investigate the EVD growth dynamics on a local scale, we analyzed subnational EVD case time series in the three most affected countries namely Guinea, Sierra Leone, and Liberia. We compiled publicly available weekly time series of reported EVD case numbers by administrative division (county or district) from the patient database available from the World Health Organization website for the period 05-Jan to 17-Dec 2014 [16]. Because some divisions have experienced more than one outbreak as evidenced by periods of non-zero case incidence separated by at least 21 consecutive days (maximum incubation period of EVD) with no new EVD reports, we focused our analysis on the time period covering the largest outbreak of EVD in each district. To avoid the effect of demographic noise in smaller outbreaks, we only considered locales that have reported at least 100 cases over the study period.

For each administrative division, we assessed whether cumulative EVD case counts could be approximated by exponential or polynomial growth during at least 3 consecutive disease generations of EVD (i.e., about 6 weeks where the mean generation time of EVD is ~15 days [12]). For this purpose, we first plotted local epidemic curves on semi-logarithmic scale. In semi-logarithmic scale, exponential growth is evident if a straight line fits well several consecutive disease generations of the epidemic curve, whereas a strong curvature in semi-logarithmic scale would be indicative of sub-exponential growth as illustrated in Figure 1. For instance, a straight line fitted to the square-root transformed epidemic curve would be indicative of quadratic polynomial growth.

To fit polynomial growth models, we borrowed from an approach used to characterize the early spread of the HIV/AIDS epidemic [17] and fit $m$-th degree polynomials through least-squares regression to the cumulative incidence over 3 or more disease generations, $C_i(t) = c_{i0} + c_{i1} t^m$. Here $C_i(t)$ is the cumulative incidence in $i$-th local district or (localized region) at time $t$, in days, since the start of the epidemic and $c_{i0}, c_{i1}$ are constants that depend specific district. Separate models are fitted to data from each district.

In the polynomial model, the relative growth rate, $dC(t)/dt \propto m/t$, decreases inversely with time while the doubling time $T_d = t (\ln 2)/m$ increases proportionally with time. This differs from the constant doubling time of epidemics characterized by exponential growth such as unmitigated outbreaks of influenza and measles [14].

**Results**

The timing of EVD outbreak across the districts in Guinea, Sierra Leone, and Liberia is shown in Figure 2. On average a new district was EVD-infected every ~ 13days (SD=14.05) in Guinea, ~10.2 days (SD=11.3) in Sierra Leone, and 13 days in Liberia (SD=15.5).

We found significant differences in the growth patterns of EVD cases at the scale of the country, district, and other subnational administrative divisions. For West Africa as a whole, the cumulative number of EVD cases initially grew exponentially, a pattern that was driven by exponential growth dynamics in Guinea where the epidemic began in December 2013 (Figure 3). The aggregate data for West Africa from early June to about mid-September 2014 can be characterized by a second exponential growth phase, albeit with a lower intrinsic growth rate than the first (Figure 3). At the national level, Liberia experienced a small wave of infections during late March to early June 2014, a pattern that was followed by exponential growth until about mid-September 2014. For Sierra Leone, exponential growth describes well the epidemic data from mid-July to late October 2014.

In contrast, analysis of subnational data reveals asynchronous transmission throughout the region. Further, local epidemics display slower, sub-exponential, growth patterns, as semi-logarithmic plots display a strong curvature during 3 or more EVD generations that can be better fitted using a quadratic polynomial rather than a straight line (Figures 4-6). Specifically, linear or quadratic growth fits well subnational EVD epidemics in Guinea (Figure 7), Sierra Leone (Figure 8), and Liberia (Figure 9) while local epidemic curves quickly depart from the expected linear trend in semi-logarithmic scale for exponentially growing epidemics. The sub-exponential growth pattern also includes the Guinean district of Gueckedou, the initial foci of the Ebola outbreak, where EVD transmission chains progressed unchecked for at least 3 months (Figure 7).

**Discussion**

We have shown that the cumulative number of EVD cases in a number of administrative areas of Guinea, Sierra Leone and Liberia is best approximated by polynomial rather than exponential growth over several generations of EVD. This pattern is also true of the initial diffusion of the West African EVD epidemic, particularly in Gueckedou district, Guinea, where a super-spreading event linked to burial rituals facilitated onward transmission in December 2013. Transmission progressed unchecked for several generations in Gueckedou, as the World Health Organization did not receive notification of the outbreak until March 23rd, 2014 [18]. In contrast, when data are aggregated nationally, or across the broader West Africa region, total case counts show periods of approximate exponential growth, in part due to an increasing number of areas affected by EVD.

Our findings raise questions on the mechanisms responsible for the slower than exponential growth rate of local Ebola outbreaks in West Africa. A review of the literature indicates that two main mechanisms could explain the approximately polynomial growth patterns: 1) the intrinsic epidemiology of the disease relating to the heterogeneities in the contact network that may "waste" infectious contacts (such as biased mixing of the population [17] or significant levels of clustering [19]), and 2) a gradually increasing impact of population behavior changes and control interventions that may decrease contacts between susceptible and infected and therefore the effective reproduction number [19-21]. Because the polynomial growth pattern characterizes even the early stages of the epidemic in the initial foci of infection in Guinea, these effects are important to consider at every stage of the epidemic.

It is well known that the characteristics of the underlying network of contacts, which a disease spreads through, play an important role on the transmission dynamics [14,22-33]. Many transmission models assuming random mixing of the population (e.g., each pair of individuals in the population has the same probability of contact) predict exponential growth, until the number of susceptible individuals in the population is significantly depleted and saturation effects blunt transmission [14,34]. At the other end of the spectrum of population mixing patterns are spatial networks where each individual has only contacts with their nearest neighbors (such as close family members), which leads to high levels of clustering [35,36] and polynomial growth.

Simulations have shown that high-levels of network clustering may be responsible for the polynomial growth of disease spread [19] because the contacts of recently infected individuals are likely to be already infected by other infectious individuals [19,37]. The spatial dissemination of the disease can be captured by a square spatial lattice covering a region, where each spatial location (node) has 4 local contacts in four directions: north, south, east and west. As expected, disease spreads in this network as a moving wave through space and the cumulative number of new cases grows as the square of time, which is the same rate of growth that we have detected in our subnational analyses of the Ebola epidemic in West Africa.

Hence, we hypothesize that the polynomial growth of EVD in West Africa could be explained by constrained transmission dynamics in a localized spatial structure of contacts shaped by the epidemiology of EVD [38]. In particular, if infections are mostly limited to people who live close to the infected population then a polynomial growth rate is to be expected. This is consistent with the assumption that infected people in the later, more severe, stages of the disease tend to be confined at home or in hospital where infectious contacts are mostly limited to local caregivers (health care workers, family members).

However, transmission is also possible during "super spreading events" involving unsafe funeral rituals where multiple mourners touch the deceased. This phenomenon likely amplified EBV transmission during the early stages of West African epidemic and was responsible for the long-distance dissemination of the disease to distant villages and other countries [2]. Also, the number of long-distance transmission events, which correspond to sporadic sparks of infection, must be relatively uncommon so that the local growth rate does not become exponential, but sufficient for regional and national epidemics to follow approximately exponential growth.

The "Small World" network model of Watts and Strogatz in 1998 [23] can be used to analyze the impact of very short distances between individual nodes, while at the same time preserving significant levels of clustering by only introducing a small number of long-distance connections in a one-dimensional lattice. However, even in this setting, an epidemic can spread exponentially fast in the network even in the presence of a very small fraction of long-distance links compared to the number of nearest-neighbor (local) links.

The slower than exponential growth of the Ebola epidemic could also be explained by an increasing effect of population behavior changes and control interventions taking hold in the population over time in such a way that the effective reproduction number quickly decreases and approaches the epidemic threshold at 1.0 [5,20,21]. However, even in the district of Gueckedou, Guinea associated with the epicenter of the epidemic, the rate of spread exhibited sub-exponential growth before the start of interventions as indicated by the strong curvature in the plot in logarithmic scale shown in Figure 4.

Our study is prone to a number of limitations. Most importantly, we rely on case counts provided by WHO, and are unable to account for reporting delays and back filling. It is unlikely however that the slow growth pattern observed in 22 administrative divisions throughout West Africa is exclusively the consequence of reporting delays. Further, we only provide here a descriptive analysis of EBV growth patterns and do not fit our data to mechanistic transmission models. Further research should concentrate on fitting EBV transmission models under different scenarios of behavior changes, interventions, and contact clustering, to the available data, to identify the most likely factors at play.

Polynomial growth has also been observed in other epidemics such as HIV/AIDS, which is spread largely through close contact via bodily fluids. The HIV/AIDS epidemic is characterized by a transmission network that deeply violate the underlying assumptions for a basic SIR model [17]. It is well documented that the cumulative number of HIV/AIDS cases in the United States in the 1980s grew as a cubic polynomial in time rather than exponentially as was initially expected by most modelers [17], and a similar mechanism could play a plausible role in the spread of EVD.

[17].

As in previous international health crises such as the SARS epidemic and the 2009 influenza pandemic, mathematical models are proving instrumental to forecast the expected total number of infections and deaths at the end of the outbreak (e.g., [6,12,13,39-43]). Previous statistical predictions based on the overall cumulative EVD cases for all Western Africa have assumed exponential growth [4-13]. Random mixing transmission model forecasts with exponential case growth greatly overestimate the final epidemic size [6,8,12,13,39]. More accurate statistical forecasts should consider the possibility of slowly growing local outbreaks that are asynchronous in time. Faster spread may be impeded by network heterogeneity (the most plausible mechanism in the early stage of the uncontrolled outbreak), helped in recent weeks by increased mitigation efforts and behavior changes of the susceptible population. These mechanisms may cause the exponential growth period to quickly end, as observed in the recent Liberian case counts [44-46]. Consequently, crude EVD epidemic forecasts assuming exponential growth should be limited to predict case burden within the immediate 2-4 generations of the disease (i.e., 1-2 months).

Our observations demonstrate how analyzing the EVD epidemic data on a local scale can reveal differences between this epidemic and more typical epidemics (such as influenza) that initially grow exponentially. These observations, together with further research on the mechanisms behind the approximately polynomial growth of local EVD outbreaks, can help develop more accurate transmission models for EVD that may be better equipped to make useful predictions.

**Competing interests**

The authors have declared that no competing interests exist.

**Financial disclosure**

The funders had no role in study design, data collection and analysis, decision to publish, or preparation of the manuscript

**Funding information**

CV and GC acknowledge the financial support from the Division of International Epidemiology and Population Studies, The Fogarty International Center, United States National Institutes of Health, funded in part by the Office of Pandemics and Emerging Threats at the United States Department of Health and Human Services. GC also acknowledges support from grants NSF grant 1414374 as part of the joint NSF-NIH-USDA Ecology and Evolution of Infectious Diseases programme, United Kingdom Biotechnology and Biological Sciences Research Council grant BB/M008894/1, NSF-IIS Grant#1518939 "RAPID: Understanding the Evolution Patterns of the Ebola Outbreak in West-Africa and Supporting Real-Time Decision Making and Hypothesis Testing through Large Scale Simulations". We would like to acknowledge the Lundbeck Foundation and the Research And Policy for Infectious Disease Dynamics program (RAPIDD) of the United States Department of Homeland Security for sponsorship of LS. JMH acknowledges partial support from NIH/NIGMS/ Modeling Infectious Disease Agent Study (MIDAS) grant U01-GM097658-01 and the NSF/DEB RAPID award B53035G1.

**Figures**

**Figure 1:**
Epidemic curves in semi-logarithmic scale illustrate the exponential growth and sub-exponential growth patterns. In semi-logarithmic scale, exponential growth is evident if a straight line fits well several consecutive disease generations of the epidemic curve whereas a strong curvature in the epidemic curve would be indicative of sub-exponential growth.

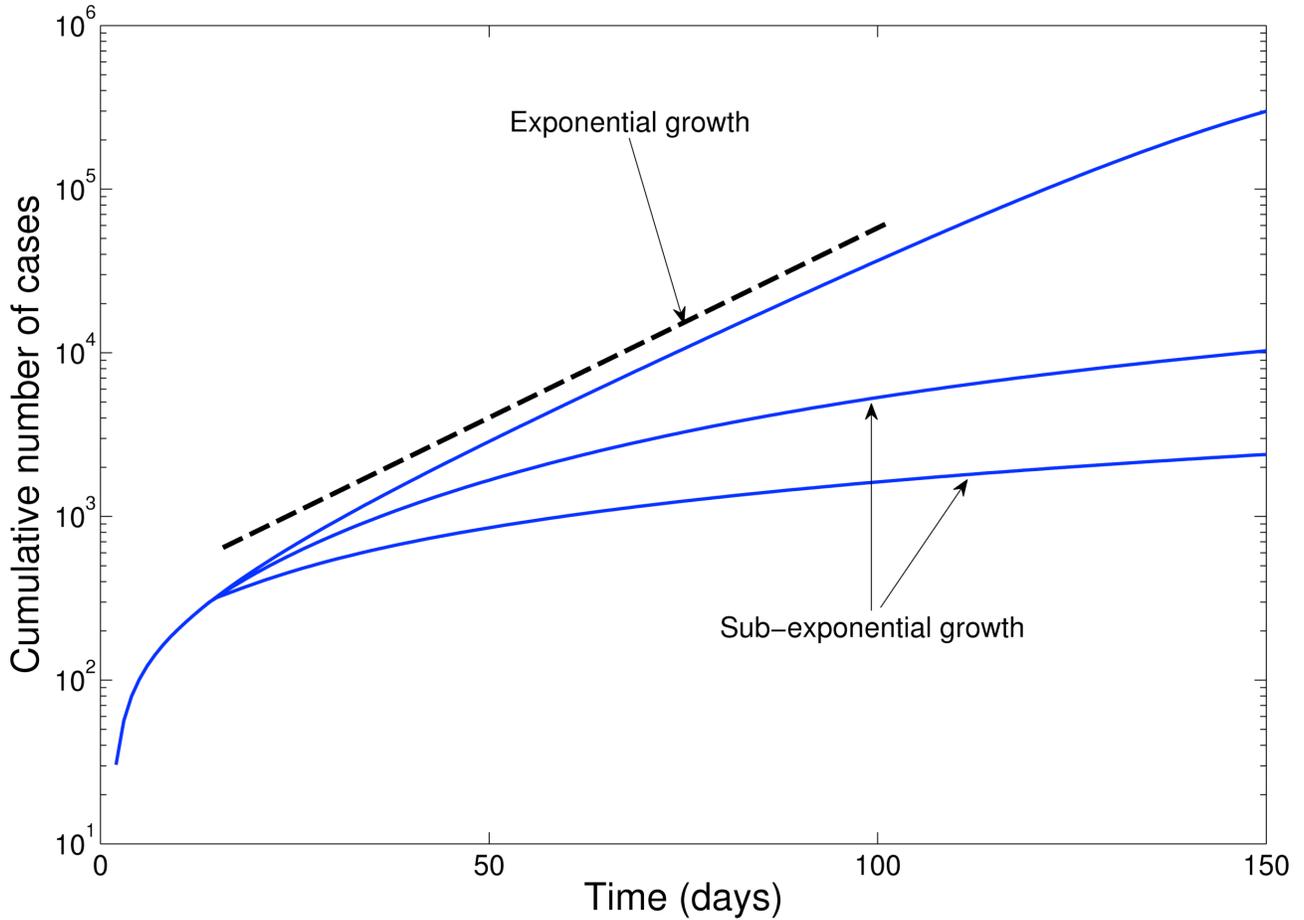

**Figure 2:**
Timing of EVD outbreak onset across districts of Guinea, Sierra Leone, and Liberia. On average a new district was EVD-infected every ~ 13days (SD=14.05) in Guinea, ~10.2 days (SD=11.3) in Sierra Leone, and 13 days in Liberia (SD=15.5).

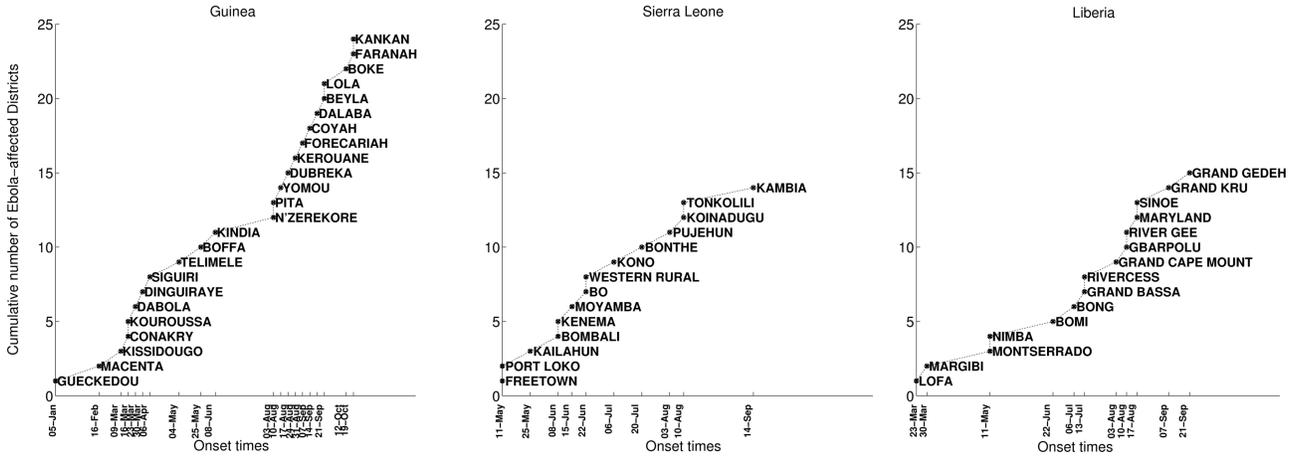

**Figure 3:**
The cumulative number of EVD cases (in log scale) in Guinea, Sierra Leone, and Liberia and the aggregate curve for all of West Africa. The cumulative number of EVD cases initially grew exponentially based on aggregate data for West Africa as a whole, a pattern that was driven by exponential growth dynamics in Guinea where the epidemic began in December 2013. The aggregate data for West Africa from early June to about mid-September 2014 can also be characterized by a second exponential growth phase, albeit with a lower intrinsic growth rate than the first. At the national level, Liberia experienced a small wave of infections during late March to early June 2014, a pattern that was followed by exponential growth until about mid-September 2014. For the aggregate data for Sierra Leone, exponential growth describes well the epidemic data from mid-July to late October 2014.

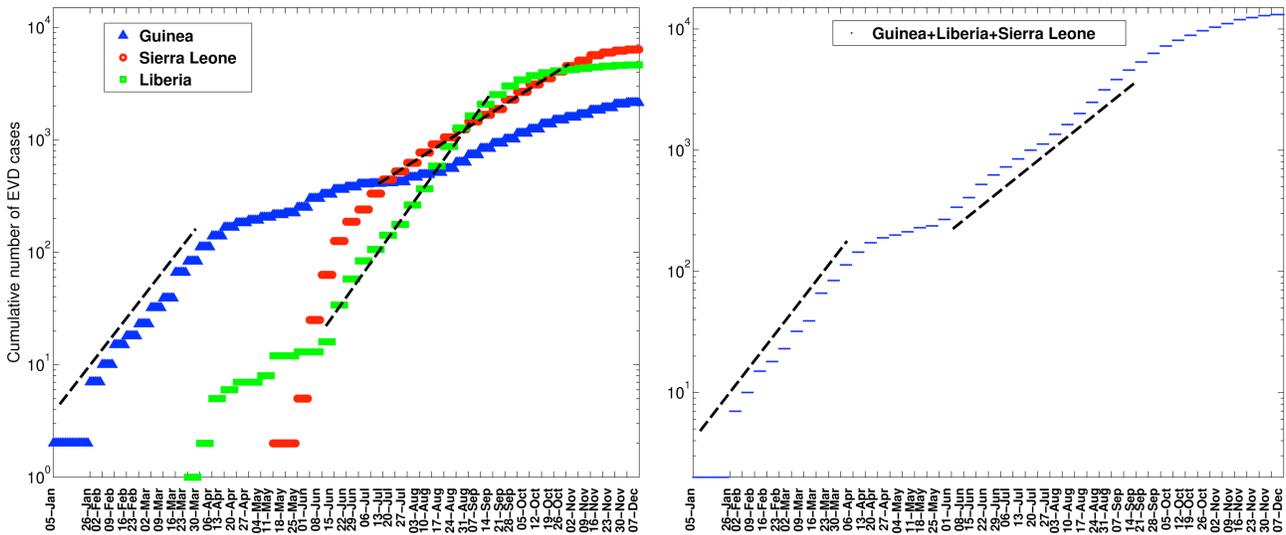

**Figure 4:**

The cumulative number of EVD cases on a semi-logarithmic plot by district and in all of Guinea. The aggregated epidemic curve for Guinea initially followed an exponential growth pattern as indicated by the initial linear trend in the log-transformed epidemic curve and highlighted by the black solid line. The district level curves are largely characterized by sub-exponential growth, shown by the strong curvature in the curves, which can be well fitted using a polynomial curve. The vertical dashed lines indicate the timing of notification of the outbreak to the World Health Organization.

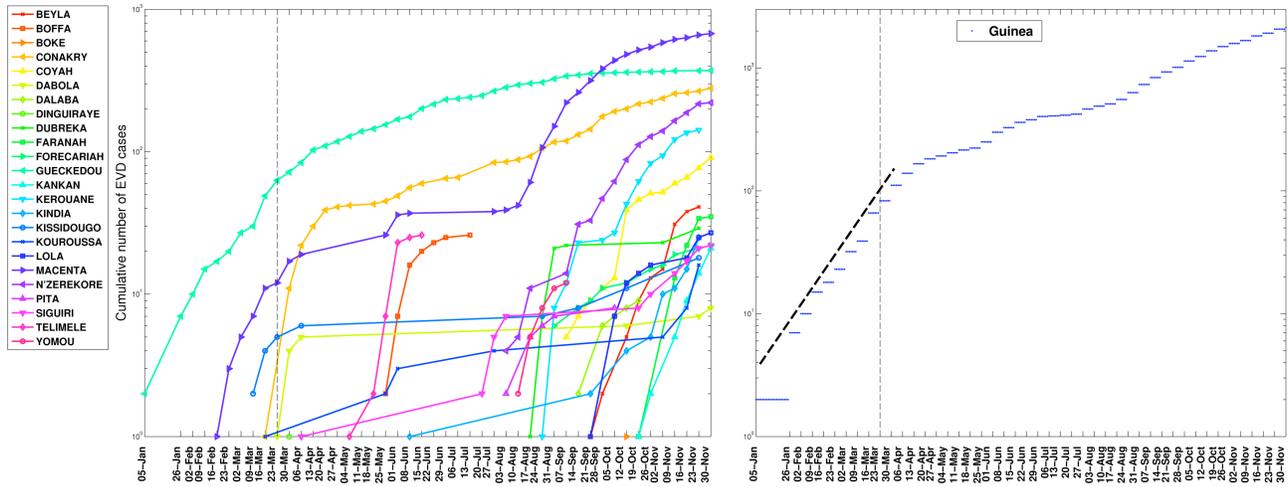

**Figure 5:**

The cumulative number of EVD cases on a semi-logarithmic plot by district and in all of Sierra Leone. The district level curves are largely characterized by sub-exponential growth, shown by the strong curvature in the curves, which can be well fitted using a polynomial curve.

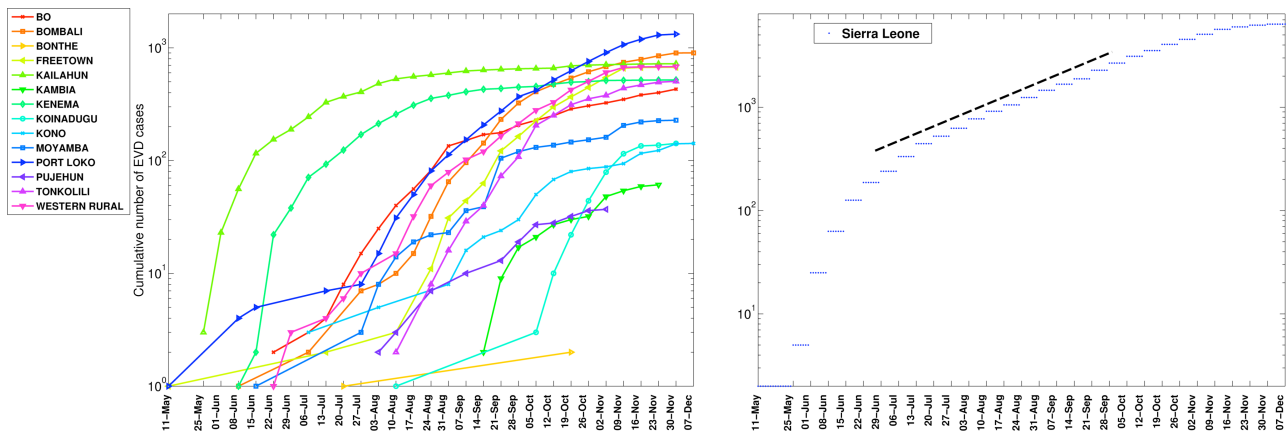

**Figure 6:**
The cumulative number of EVD cases on a semi-logarithmic plot by district and in all of Liberia. The district level curves are largely characterized by sub-exponential growth, shown by the strong curvature in the curves, which can be well fitted using a polynomial curve.

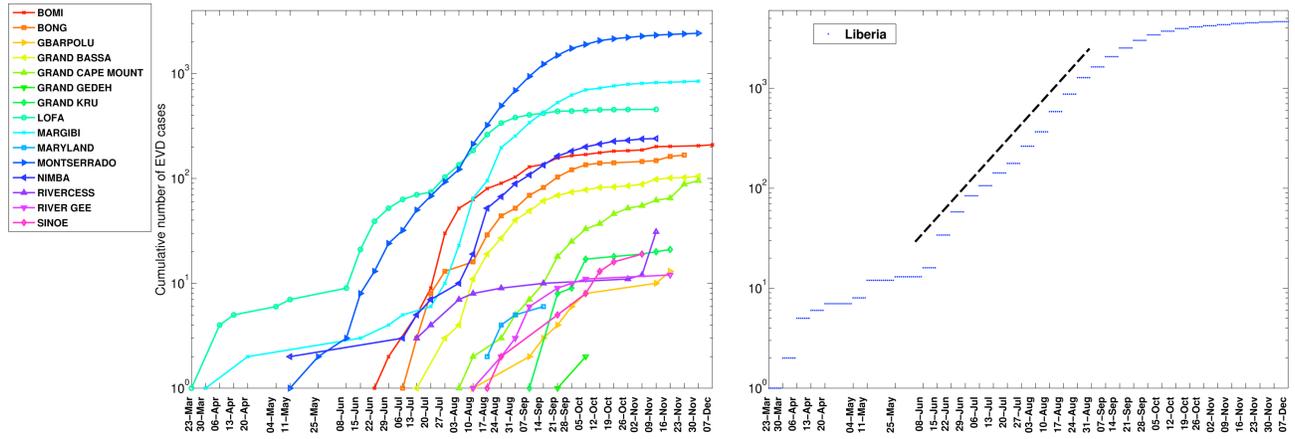

**Figure 7:**
Cumulative number of EVD cases in districts of Guinea reporting over 100 cases. Three different transformations of the epidemic data are shown: 1) raw data (no transformation, left y-scale), 2) log-transformation (right y-scale), and 3) square-root transformation (right y-scale). Epidemic curves show slower, sub-exponential, growth patterns, as semi-logarithmic plots display a strong curvature during 3 or more EVD generations that can be better fitted by linear or quadratic growth. Local epidemic curves quickly depart from the expected linear trend in log-transformed data. The dashed line, shown as a reference, was fitted to the first 4 weeks of the log-transformed data.

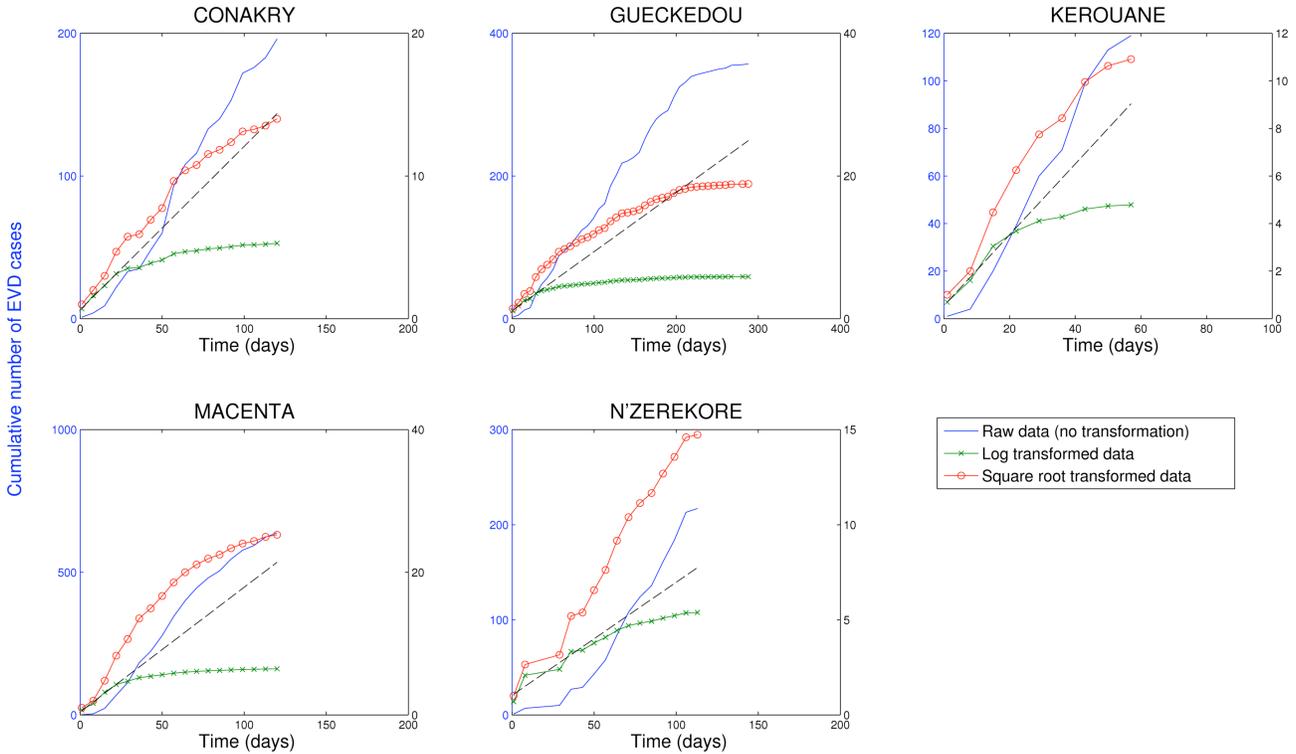

**Figure 8:**
Cumulative number of EVD cases in districts of Sierra Leone reporting over 100 cases. Three different transformations of the epidemic data are shown: 1) raw data (no transformation, left y-scale), 2) log-transformation (right y-scale), and 3) square-root transformation (right y-scale). Epidemic curves show slower, sub-exponential, growth patterns, as semi-logarithmic plots display a strong curvature during 3 or more EVD generations that can be better fitted by linear or quadratic growth. Local epidemic curves quickly depart from the expected linear trend in log-transformed data. The dashed line, shown as a reference, was fitted to the first 4 weeks of the log-transformed data.

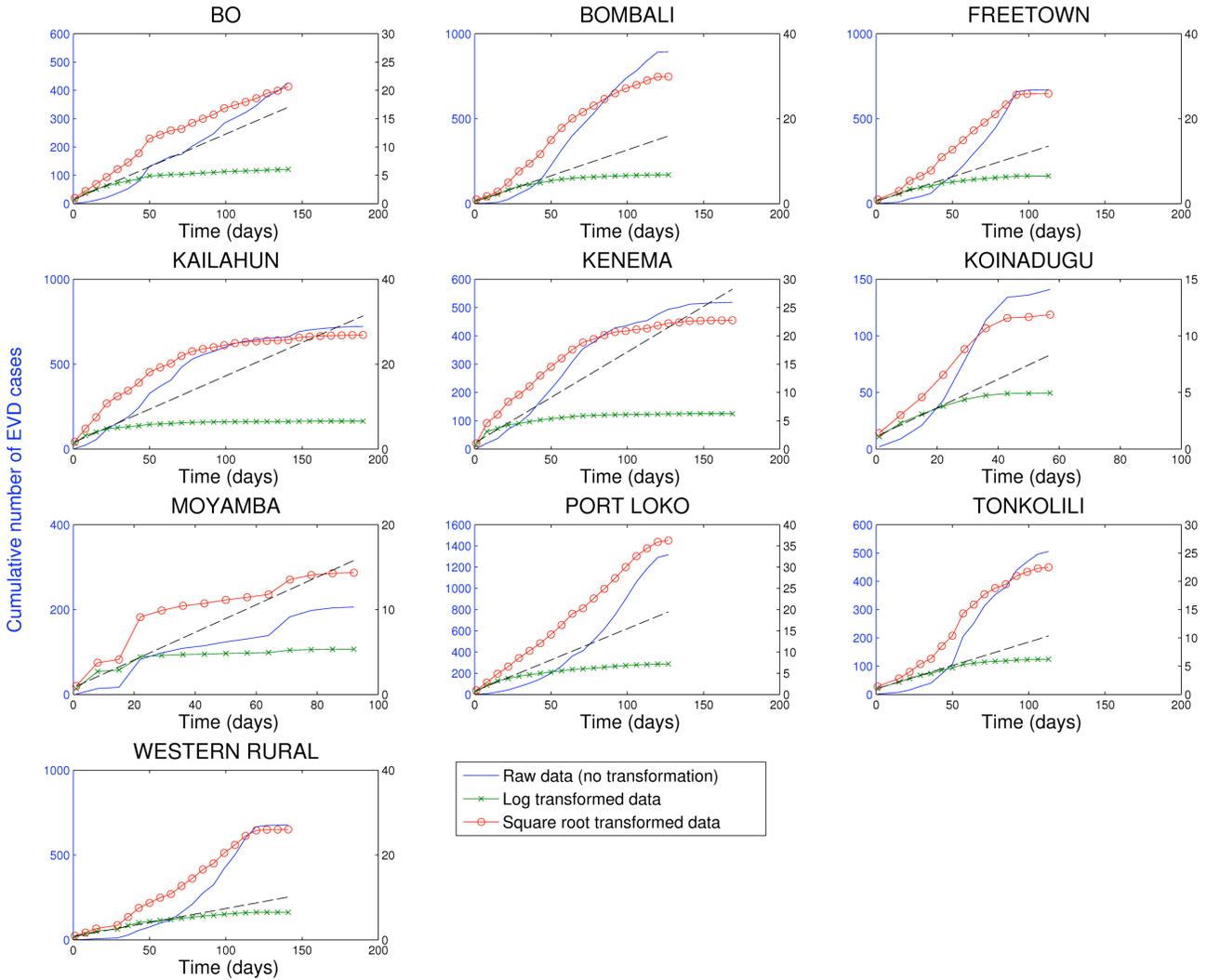

**Figure 9:**
Cumulative number of EVD cases in districts of Liberia reporting over 100 cases. Three different transformations of the epidemic data are shown: 1) raw data (no transformation, left y-scale), 2) log-transformation (right y-scale), and 3) square-root transformation (right y-scale). Epidemic curves show slower, sub-exponential, growth patterns, as semi-logarithmic plots display a strong curvature during 3 or more EVD generations that can be better fitted by linear or quadratic growth. Local epidemic curves quickly depart from the expected linear trend in log-transformed data. The dashed line, shown as a reference, was fitted to the first 4 weeks of the log-transformed data.

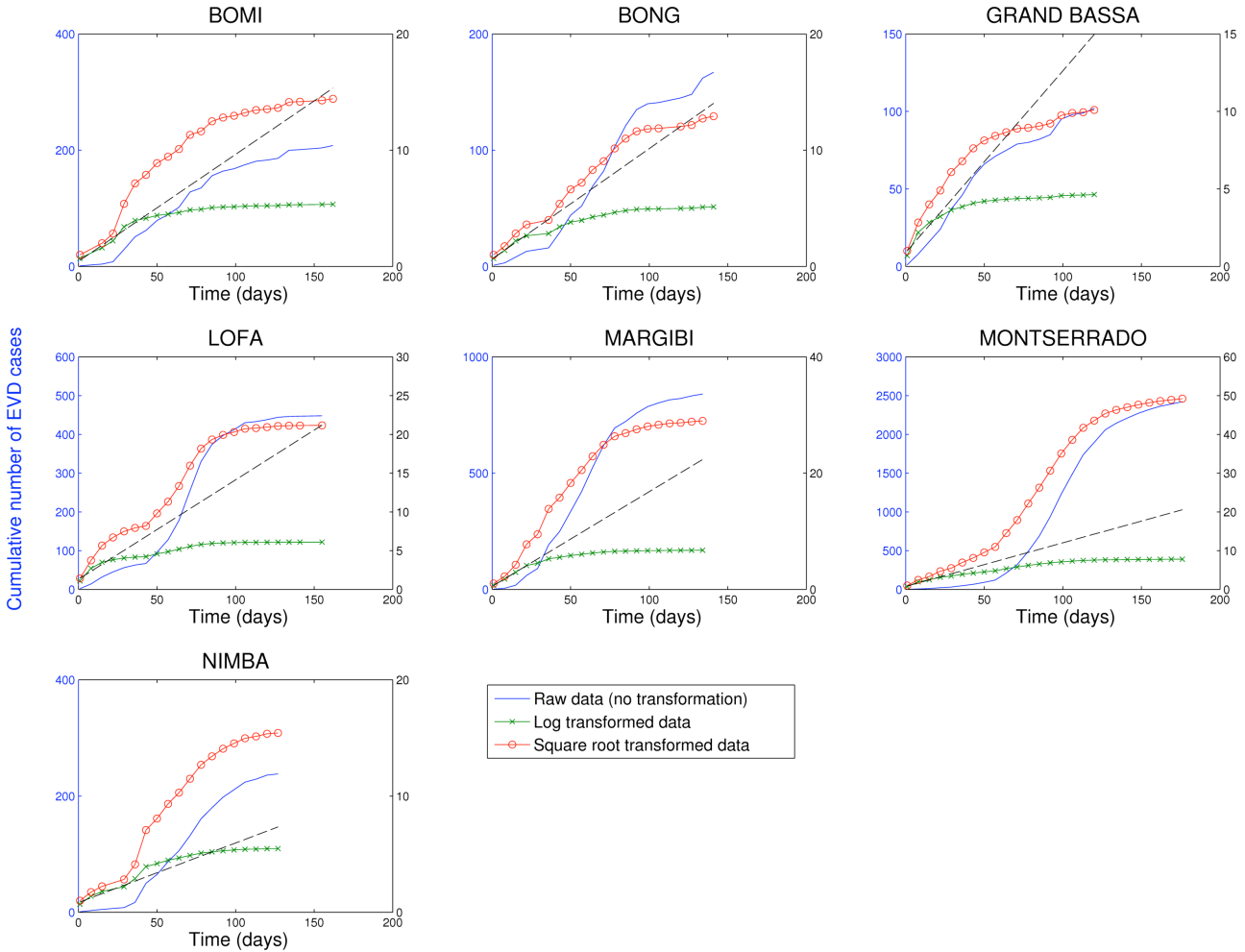